\documentclass[english,aps,nofootinbib,twocolumn,preprintnumbers,nobalancelastpage,floatfix,superscriptaddress,groupedaddress,prl,reprint]{revtex4-1}

\usepackage[latin1]{inputenc}
\pdfoutput = 1
\usepackage{bm}
\usepackage{color}
\usepackage{graphicx}
\usepackage{cancel}
\usepackage{amsfonts}
\usepackage{amssymb}
\usepackage{amsmath}
\usepackage{calc}
\usepackage{dcolumn}
\usepackage{mathrsfs}

\usepackage{soul}

\usepackage[normalem]{ulem}             

\newcommand{\eg}{e.g.~}
\newcommand{\ie}{i.e.~}

\newcommand{\Eq}[1]{Eq.~\eqref{#1}}
\newcommand{\Fig}[1]{Fig.~\ref{#1}}

\newcommand{\Lag}{\mathscr{L}}	

\newcommand{\beq}{\begin{equation}}
\newcommand{\eeq}{\end{equation}}

\newcommand{\ud}{\text{d}}
\newcommand{\bol}[1]{\boldsymbol{#1}}

\newcommand{\ER}{E_\text{R}}

\newcommand{\vmin}{v_\text{min}}

\newcommand{\tmax}{t_\text{max}}
\newcommand{\tmin}{t_\text{min}}
\newcommand{\htmin}{\hat{t}_\text{min}}
\newcommand{\taumax}{\tau_\text{max}}
\newcommand{\taumin}{\tau_\text{min}}
\newcommand{\htaumin}{\hat{\tau}_\text{min}}

\definecolor{rossoCP3}{cmyk}{0,.88,.77,.40}
\definecolor{verdeCP3}{rgb}{0.09765625, 0.57421875, 0.1015625}
\definecolor{bluCP3}{rgb}{0, 0.23, 0.67}

\ifx\pdfoutput\undefined
\usepackage[dvips,bookmarks]{hyperref}    
\else
\usepackage{hyperref}    
\fi
\hypersetup{colorlinks, bookmarksopen, bookmarksnumbered,
citecolor=verdeCP3, linkcolor=bluCP3, pdfstartview=FitH, urlcolor=rossoCP3}

\newcommand{\AddrUCLA}{Department of Physics and Astronomy, UCLA, 475 Portola Plaza, Los Angeles, CA 90095 (USA)}

\begin{document}

\title{Target dependence of the annual modulation in direct dark matter searches}

\author{Eugenio Del Nobile}
\author{Graciela B.~Gelmini}
\author{Samuel J.~Witte}

\affiliation{\AddrUCLA}

\begin{abstract}
Due to Earth's revolution around the Sun, the expected scattering rate in direct dark matter searches is annually modulated. This modulation is expected to differ between experiments when given as a function of recoil energy $\ER$, \eg due to the gravitational focusing effect of the Sun. A better variable to compare results among experiments employing different targets is the minimum speed $\vmin$ a dark matter particle must have to impart a recoil energy $\ER$ to a target nucleus. It is widely believed that the modulation expressed as a function of $\vmin$ is common to all experiments, irrespective of the dark matter distribution. We point out that the annual modulation as a function of $\vmin$, and in particular the times at which the rate is maximum and minimum, could be very different depending on the detector material. This would be an indication of a scattering cross section with non-factorizable velocity and target material dependence. Observing an annual modulation with at least two different target elements would be necessary to identify this type of cross section.
\end{abstract}

\maketitle


\section{Introduction}

Dark matter (DM) is the most abundant form of matter in the Universe and its nature still remains a mystery. More than $80\%$ of the mass of our galaxy resides in a spheroidal DM halo, which extends well beyond the visible disk. Efforts to detect new elementary particles which could constitute the DM are multi-pronged.

Direct DM detection experiments attempt to detect the energy deposited by DM particles in the dark halo of our galaxy when they collide with nuclei inside a detector. An unmistakable signature of the expected DM signal is an annual modulation of the rate caused by the rotation of Earth around the Sun \cite{Drukier:1986tm}. For DM velocity distributions that are smooth and isotropic in the galactic frame at Earth's location, the expected differential rate for DM scattering onto a target nuclide $T$ in all direct DM detection experiments could be well represented by the first two terms of a harmonic expansion (see \eg \cite{Lee:2013xxa}),
\beq\label{modrate}
\frac{\ud R_T}{\ud \ER}(\ER, t) = S_0(\ER) + S_\text{m}(\ER) \cos \! \left( \frac{2 \pi}{1 \text{ year}} (t - t_0) \right) .
\eeq
Here $\ER$ is the nuclear recoil energy and $t_0$ is the time at which the speed of Earth with respect to the galaxy is maximum, close to June $1^\text{st}$. At high $\ER$, with $S_\text{m}$ positive $t_0$ equals the time $\tmax$ at which the rate is maximum, while $\tmin$, the time at which the rate is minimum, is six months apart from $\tmax$ (except for a shift of about a day due to the eccentricity of Earth's orbit). At low $\ER$, $S_\text{m}$ could become negative, implying $t_0$ equals $\tmin$ instead of $\tmax$ (see \eg Fig.~8.2 of \cite{Smith:1988kw}). Anisotropies in the local DM velocity distribution modify this picture, in particular by making $\tmax$ and $\tmin$ energy dependent. The gravitational focusing (GF) of DM particles due to the Sun inherently makes the local DM halo anisotropic \cite{Alenazi:2006wu}. Ref.~\cite{Lee:2013wza} has shown GF to have a significant effect on the phase of the modulation at low enough recoil energy.

Since $\ER$ depends on the target nuclide mass, it is not a good variable to compare the annual modulation of the rate among experiments employing different targets. A better variable is $\vmin$, the minimum speed a DM particle must have in Earth's rest frame to impart a recoil energy $\ER$ onto a target nucleus. It is typically assumed that $\tmax$ and $\tmin$ as functions of $\vmin$ do not depend on the target, and consequently they can be used to test the agreement between putative DM signals across multiple detectors.

Here we point out that, in general, the annual modulation of the rate as a function of $\vmin$ can vary significantly for different target materials. Specifically, we show that if the velocity and target dependence cannot be factored in the differential scattering cross section, observables associated with the modulation, such as $\tmax$ and $\tmin$, may be highly target dependent. Our observation does not rely on any assumption regarding the DM distribution. As an illustration, we show that for DM particles with a magnetic dipole moment $\tmax$ and $\tmin$ depend on the target material.

\section{DM signal and its modulation}

For the spin-independent and spin-dependent contact interactions usually considered, the differential scattering cross section is
\beq\label{diffsigmastandard}
\frac{\ud \sigma_T}{\ud \ER}(\ER, v) = \frac{m_T \sigma_T F_T(\ER)^2}{2 \mu_T^2} \frac{1}{v^2} \ ,
\eeq
with $m_T$ the target nuclide mass, $\mu_T$ the DM-nucleus reduced mass, $\sigma_T$ the total cross section for a point-like nucleus, and $F_T(\ER)$ the appropriate nuclear form factor. The differential scattering rate per unit target mass,
\beq\label{diffrate}
\frac{\ud R_T}{\ud \ER}(\ER, t) = \frac{C_T}{m_T} \frac{\rho}{m} \int_{v \geqslant \vmin(\ER)} v \, f(\bol{v}, t) \, \frac{\ud \sigma_T}{\ud \ER} \, \ud^3 v \ ,
\eeq
with \Eq{diffsigmastandard} becomes
\beq\label{diffratestandard}
\frac{\ud R_T}{\ud \ER}(\ER, t) = C_T \frac{\rho}{m} \frac{\sigma_T F_T(\ER)^2}{2 \mu_T^2} \, \eta(\vmin(\ER), t) \ ,
\eeq
with $\rho$ and $m$ the local DM particle density and mass, respectively, and $C_T$ the nuclide mass fraction in the detector. Here we defined the velocity integral
\beq\label{standardeta}
\eta(\vmin, t) \equiv \int_{v \geqslant \vmin} \frac{f(\bol{v}, t)}{v} \, \ud^3 v \ ,
\eeq
where $f(\bol{v}, t)$ is the DM velocity distribution in Earth's frame. The time dependence arises due to Earth's revolution around the Sun. The modulation of the rate in \Eq{diffratestandard} is determined by the time dependence of $\eta(\vmin, t)$, which is common to all experiments. Therefore, for the interaction in \Eq{diffsigmastandard}, $\tmax$ and $\tmin$ for fixed $\vmin$ do not depend on the target material. This remains true for other differential cross sections where the velocity and target dependences can be factored. In general, however, the differential cross section can consist of multiple terms with different velocity dependences and target-dependent coefficients, \eg with DM particles interacting through a magnetic dipole \cite{Pospelov:2000bq, Sigurdson:2004zp, Barger:2010gv, Chang:2010en, Cho:2010br, Heo:2009vt, Gardner:2008yn, Masso:2009mu, Banks:2010eh, Fortin:2011hv, An:2010kc, Kumar:2011iy, Barger:2012pf, DelNobile:2012tx, Cline:2012is, Weiner:2012cb, Tulin:2012uq, Cline:2012bz, Heo:2012dk, DelNobile:2013cva, Barello:2014uda, Lopes:2013xua, Keung:2010tu, Gresham:2013mua, DelNobile:2014eta, Gresham:2014vja} or an anapole moment \cite{Pospelov:2000bq, Ho:2012bg, Fitzpatrick:2010br, Keung:2010tu, Frandsen:2013cna, Gresham:2013mua, Frandsen:2013cna, Gao:2013vfa, DelNobile:2014eta, Gresham:2014vja}. It also happens with some of the interactions described by the effective operators studied \eg in \cite{Goodman:2010ku, Goodman:2010yf, Zheng:2010js, Liang:2013dsa, Catena:2014uqa, Catena:2014epa} (see \cite{Fitzpatrick:2012ix, Fitzpatrick:2012ib, Kumar:2013iva, DelNobile:2013sia, Barello:2014uda} for explicit formulas of scattering amplitudes). In this case the annual modulation of the rate can be strongly target element dependent.

\section{An example: magnetic dipole DM}

Here we study in detail the case of a Dirac fermion DM candidate $\chi$ that interacts with nuclei through a magnetic dipole moment $\lambda_\chi$, with interaction Lagrangian $\Lag = (\lambda_\chi / 2) \, \bar\chi \sigma_{\mu \nu} \chi F^{\mu\nu}$. The differential cross section for elastic scattering off a target nucleus $T$ with $Z_T$ protons and spin $S_T$ is
\begin{widetext}
\beq
\label{diffsigmamagnetic}
\frac{\ud \sigma_T}{\ud \ER}(\vmin, v) =
\alpha \lambda_\chi^2 \left\{ Z_T^2 \frac{m_T}{2 \mu_T^2} \left[ \frac{1}{\vmin^2} - \frac{1}{v^2} \left( 1 - \frac{\mu_T^2}{m^2} \right) \right] F_{\text{SI}, T}^2(\ER(\vmin)) + \frac{\hat\lambda_T^2}{v^2} \frac{m_T}{m_p^2} \left( \frac{S_T + 1}{3 S_T} \right) F_{\text{M}, T}^2(\ER(\vmin)) \right\} ,
\eeq
\end{widetext}
with $\alpha = e^2 / 4 \pi$ the electromagnetic constant, $m_p$ the proton mass, $\hat\lambda_T$ the nuclear magnetic moment in units of the nuclear magneton $e / (2 m_p) = 0.16$ GeV$^{-1}$, and $\ER(\vmin) = 2 \mu_T^2 \vmin^2 / m_T$. The first term is due to DM dipole-nuclear charge interaction, and the corresponding  charge form factor coincides with the usual spin-independent nuclear form factor $F_{\text{SI}, T}(\ER)$, while the second term is due to the dipole-dipole interaction and has a nuclear magnetic form factor $F_{\text{M}, T}(\ER)$ (both form factors are normalized to $1$ at zero momentum transfer). We compute the cross section with the formalism and the form factors provided in \cite{Fitzpatrick:2012ix, Fitzpatrick:2012ib}.

\begin{figure}
\centering
\includegraphics[width=0.40\textwidth, trim=1mm 3mm 1.5mm 2mm, clip]{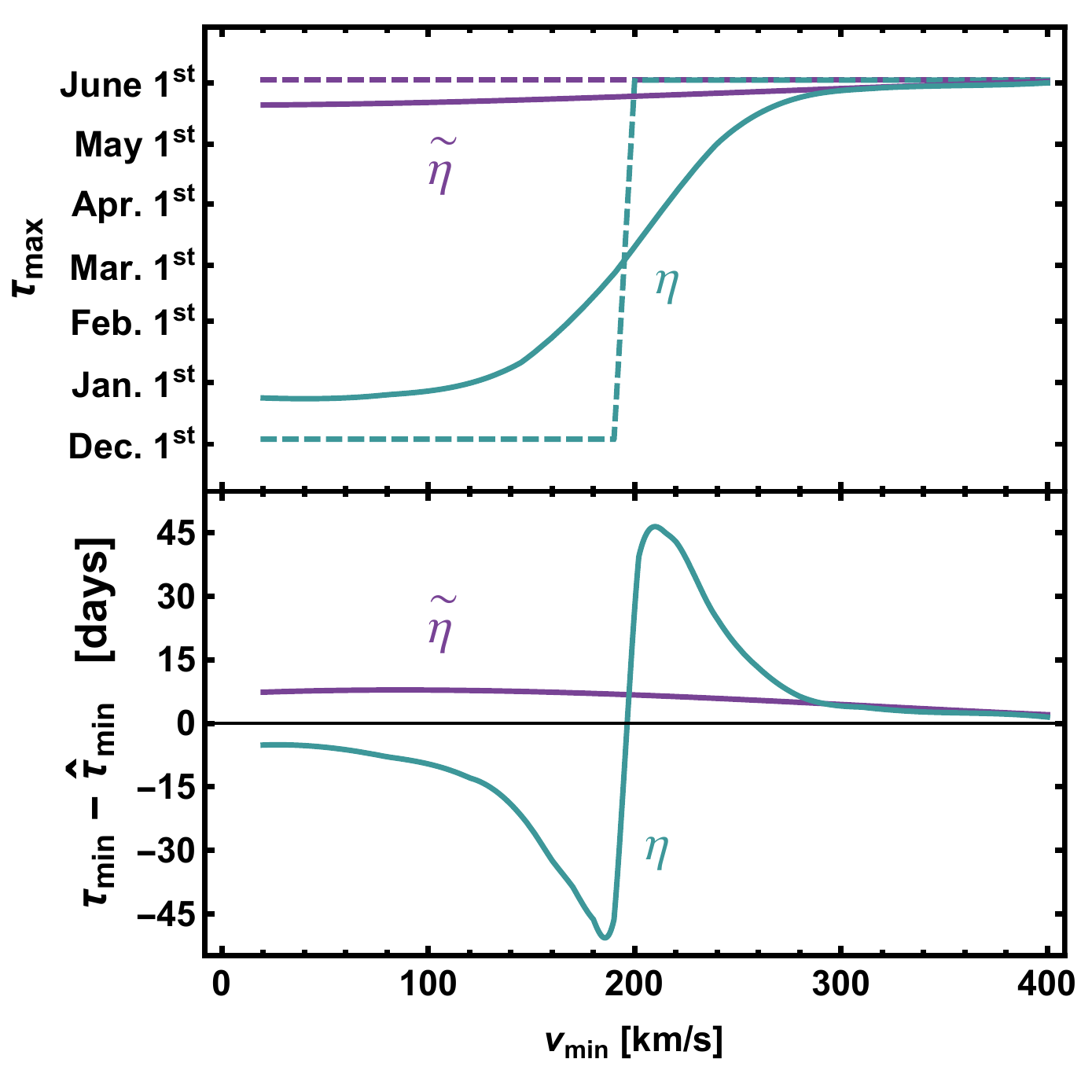}
\caption{\label{fig:eta} Time of maximum $\taumax$ ({\it top}) and minimum $\taumin$ ({\it bottom}) of $\eta$ and $\tilde{\eta}$ in the SHM, as functions of $\vmin$, including (solid lines) and neglecting (dashed lines) GF. The bottom panel shows $\taumin - \htaumin$, with $\htaumin$ the time six month apart from $\taumax$. Neglecting GF, $\taumin$ is almost indistinguishable from $\htaumin$, and thus is not shown.}
\end{figure}

The differential cross section in \Eq{diffsigmamagnetic} contains two terms with different velocity dependence: one with the usual $1 / v^2$ factor and another independent of $v$. The differential rate (see \Eq{diffrate}) is thus also a sum of two terms, one containing $\eta(\vmin, t)$ in \Eq{standardeta} and the other containing
\beq
\tilde{\eta}(\vmin, t) \equiv \int_{v \geqslant \vmin} v \, f(\bol{v}, t) \, \ud^3 v \ .
\eeq
For purposes of illustration we assume the Standard Halo Model (SHM), in which the DM velocity distribution is an isotropic Maxwellian on average at rest with respect to the galaxy (see \eg \cite{Gelmini:2014psa} for details). Under this assumption the two velocity integrals $\eta$ and $\tilde{\eta}$ have a very different time dependence. This can be seen in \Fig{fig:eta} where their time of maximum $\taumax$ and minimum $\taumin$ are shown. Instead of $\taumin$, we plot $\taumin - \htaumin$ where $\htaumin$ is the time six months apart from $\taumax$. \Fig{fig:eta} shows the effect of including (solid lines) and neglecting (dashed lines) GF. Neglecting GF, $\taumin$ is almost indistinguishable from $\htaumin$, and thus is not shown. Unless otherwise stated, we include GF and the eccentricity of Earth's orbit in our calculations. Notice that $\taumax$ ($\taumin$) as a function of $\vmin$ coincides with the maximum (minimum) of the differential rate, $\tmax$ ($\tmin$), only when the velocity and target dependence can be factored in the differential scattering cross section.

\begin{figure}[t!]
\centering
\includegraphics[width=0.43\textwidth, trim=1mm 1mm 13mm 13mm, clip]{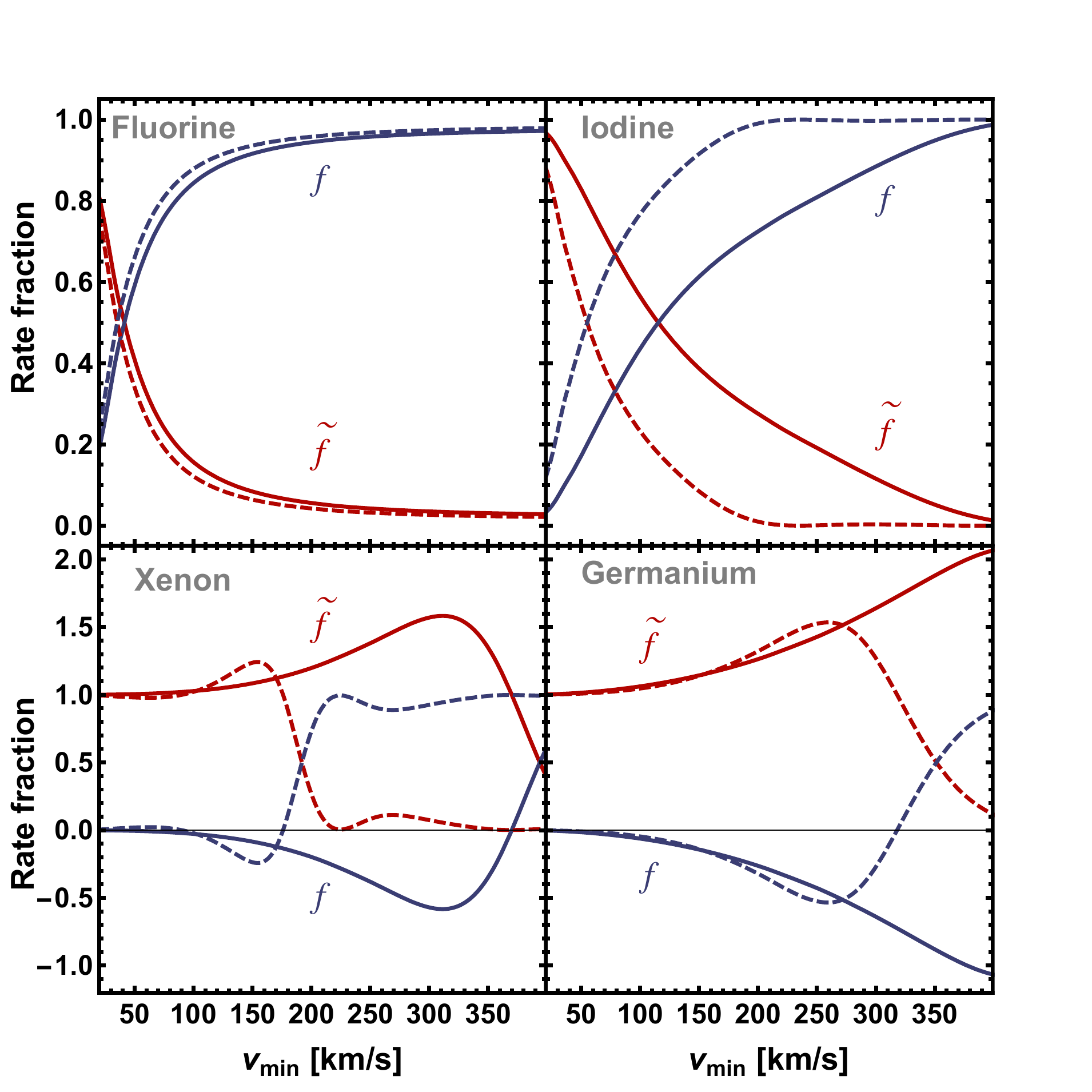}
\caption{\label{fig:ratefrac} Rate fractions $f \equiv r / (r + \tilde{r})$ and $\tilde{f} \equiv \tilde{r} / (r + \tilde{r})$ for fluorine, iodine, xenon, and germanium. Solid (dashed) lines for $m = 100$ GeV ($1$ TeV).}
\end{figure}

The modulation of the differential rate depends on the interplay of the terms containing $\eta$ and $\tilde{\eta}$. Since the relative coefficients are in general target dependent, as well as DM particle mass dependent, the modulation also depends on the target and on $m$. Let us denote with $r$ and $\tilde{r}$ the terms of the expected differential rate containing $\eta$ and $\tilde{\eta}$, so that $\ud R_T / \ud \ER = r + \tilde{r}$. \Fig{fig:ratefrac} shows the rate fractions $f \equiv r / (r + \tilde{r})$ and $\tilde{f} \equiv \tilde{r} / (r + \tilde{r})$ as functions of $\vmin$ for four different target elements (fluorine, iodine, xenon, and germanium) employed by current DM direct detection experiments. For target elements with more than one isotope (Xe, Ge), we sum \Eq{diffrate} over isotopic composition. Solid (dashed) lines in \Fig{fig:ratefrac} correspond to a $100$ GeV ($1$ TeV) DM particle. Notice that because of the negative sign in one of the dipole-charge terms in \Eq{diffsigmamagnetic}, $r$ and $f$ are allowed to take negative values. When this happens, $\tilde{f} > 1$ since $f + \tilde{f} = 1$.

\begin{figure*}[t!]
\centering
\hspace*{\fill}
\includegraphics[width=0.40\textwidth]{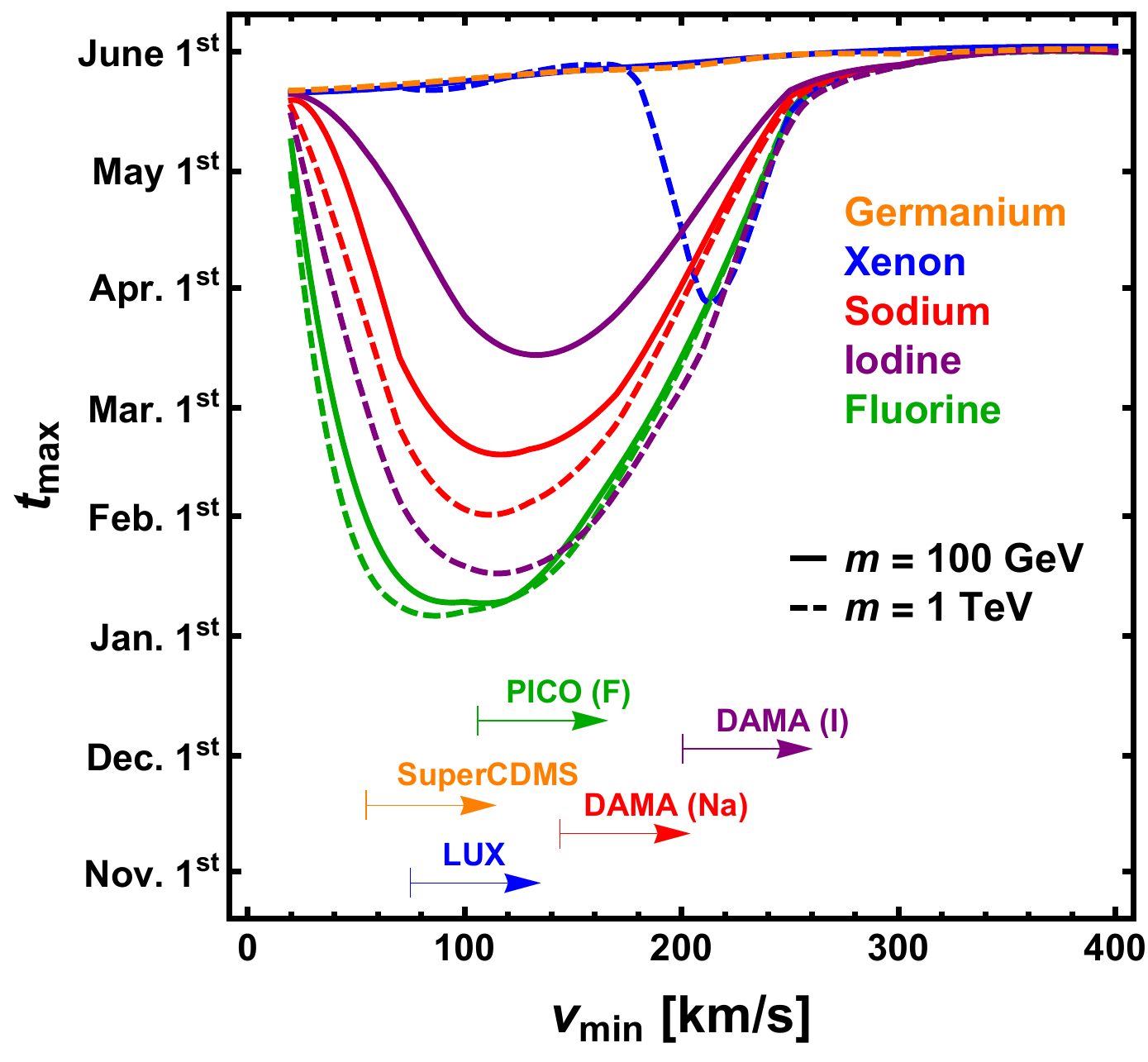}
\hfill
\includegraphics[width=0.40\textwidth]{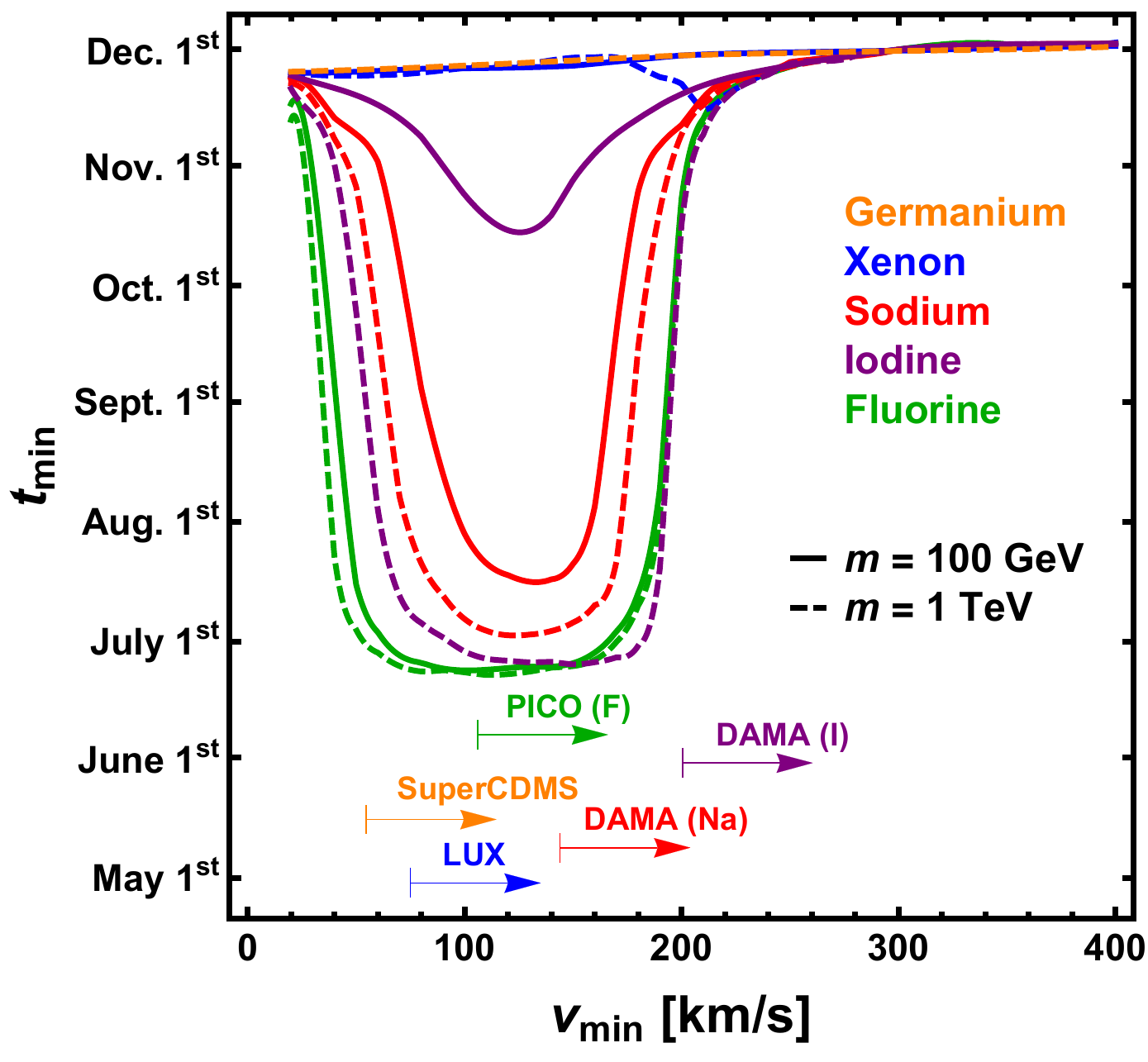}
\hspace*{\fill}
\caption{\label{fig:magneticDM} Time of maximum $\tmax$ ({\it left}) and minimum $\tmin$ ({\it right}) of the scattering rate as functions of $\vmin$ for a $100$ GeV (solid lines) and $1$ TeV (dashed lines) magnetic dipole DM particle scattering elastically off different targets, assuming the SHM. For germanium, the two lines would overlap (only dashed is shown). Also shown are the low energy thresholds for several current direct detection experiments, for $m = 100$ GeV (for larger $m$ the thresholds shift to lower $\vmin$ values).}
\end{figure*}

Figs.~\ref{fig:eta} and \ref{fig:ratefrac} can be used in combination to understand the target-dependent behavior of the time of maximum $\tmax$ and minimum $\tmin$ of the rate for magnetic DM, shown in \Fig{fig:magneticDM} for scattering off fluorine, sodium, iodine, xenon and germanium. Solid (dashed) lines correspond to $m = 100$ GeV ($1$ TeV). Also shown in \Fig{fig:magneticDM} are the $\ER$ thresholds for LUX \cite{Akerib:2013tjd} ($3.1$ keV, employing Xe), SuperCDMS \cite{Agnese:2014aze} ($1.6$ keV, Ge), DAMA \cite{Bernabei:2013xsa} ($6.7$ keV for Na and $22.2$ keV for I), and PICO \cite{Amole:2015lsj} ($3.2$ keV, F), translated into $\vmin$ for $m_T$ averaged over isotopic composition and elastic scattering for $m = 100$ GeV. For larger $m$, these thresholds move to lower $\vmin$ values.

\Fig{fig:magneticDM} shows that $\tmax$ and $\tmin$ become essentially target independent above $\vmin \simeq 300$ km/s. This is due to the fact that the differences between $\eta$ and $\tilde{\eta}$, which are central to the target dependence of the rate, rapidly vanish at $\vmin \gtrsim 300$ km/s (see \Fig{fig:eta}). The target-independent nature of this region is not specific to magnetic DM and occurs whenever the SHM is assumed, at least with $1 / v^2$ and $v^n$-dependent terms in the differential cross section and $n \geqslant 0$. This is because all velocity integrals arising from terms going as $v^{n}$ with $n \geqslant 0$ in the differential cross section have very similar phases at all $\vmin$ values, \ie they are all comparable to $\tilde{\eta}$ in \Fig{fig:eta}. The target-dependent effects addressed in this paper thus rely on having both a $1 / v^2$ term and a $v^n$ term, $n \geqslant 0$, in the differential cross section.

At sufficiently small values of $\vmin$ the rate is always dominated by $\tilde{r}$ (\ie $\tilde{f} \simeq 1$ and $f \simeq 0$), as shown in \Fig{fig:ratefrac}. This is due to the $1 / \vmin^2$ factor appearing in \Eq{diffsigmamagnetic}. Therefore in the small $\vmin$ limit one can disregard the contribution of $r$ and correctly assume $\tmax$ and $\tmin$ coincide with the $\taumax$ and $\taumin$ of $\tilde{\eta}$ shown in \Fig{fig:eta}. This explains why $\tmax$ in \Fig{fig:magneticDM} occurs in May at small $\vmin$ values regardless of the target.

Assuming at least one target isotope has a non-zero nuclear magnetic moment, the dipole-dipole part of the interaction becomes dominant, and thus $r > \tilde{r}$, at large values of $\vmin$. This is due to the fact that the spin-independent charge form factor decreases faster than the magnetic form factor. \Fig{fig:ratefrac} confirms that for the elements and DM masses considered, there is a $\vmin$ value above which $r$ dominates and below which $\tilde{r}$ dominates. In \Fig{fig:magneticDM} this corresponds to the time variation of the rate being determined by $\eta$ or $\tilde{\eta}$, respectively. For germanium, this switch occurs at large $\vmin$ values because of its small average magnetic moment. How and where this switch in $\vmin$ occurs determine the main features of $\tmax$ and $\tmin$ in \Fig{fig:magneticDM}.

For each element, the features in \Fig{fig:ratefrac} move to smaller $\vmin$ values as the DM particle mass increases. This is in part because the $\vmin$ value corresponding to a particular $\ER$ decreases, but also because the $1 / \mu_T^2$ and $\mu_T^2 / m^2$ factors in \Eq{diffsigmamagnetic} decrease. Notice that, as $m$ increases, the $\vmin$ value above which $r$ becomes the dominant term in the rate may fall below $300$ km/s, leading to the appearance of a feature in \Fig{fig:magneticDM}. This happens with xenon when $m$ goes from $100$ GeV to $1$ TeV.

\begin{figure}[t!]
\centering
\includegraphics[width=0.40\textwidth]{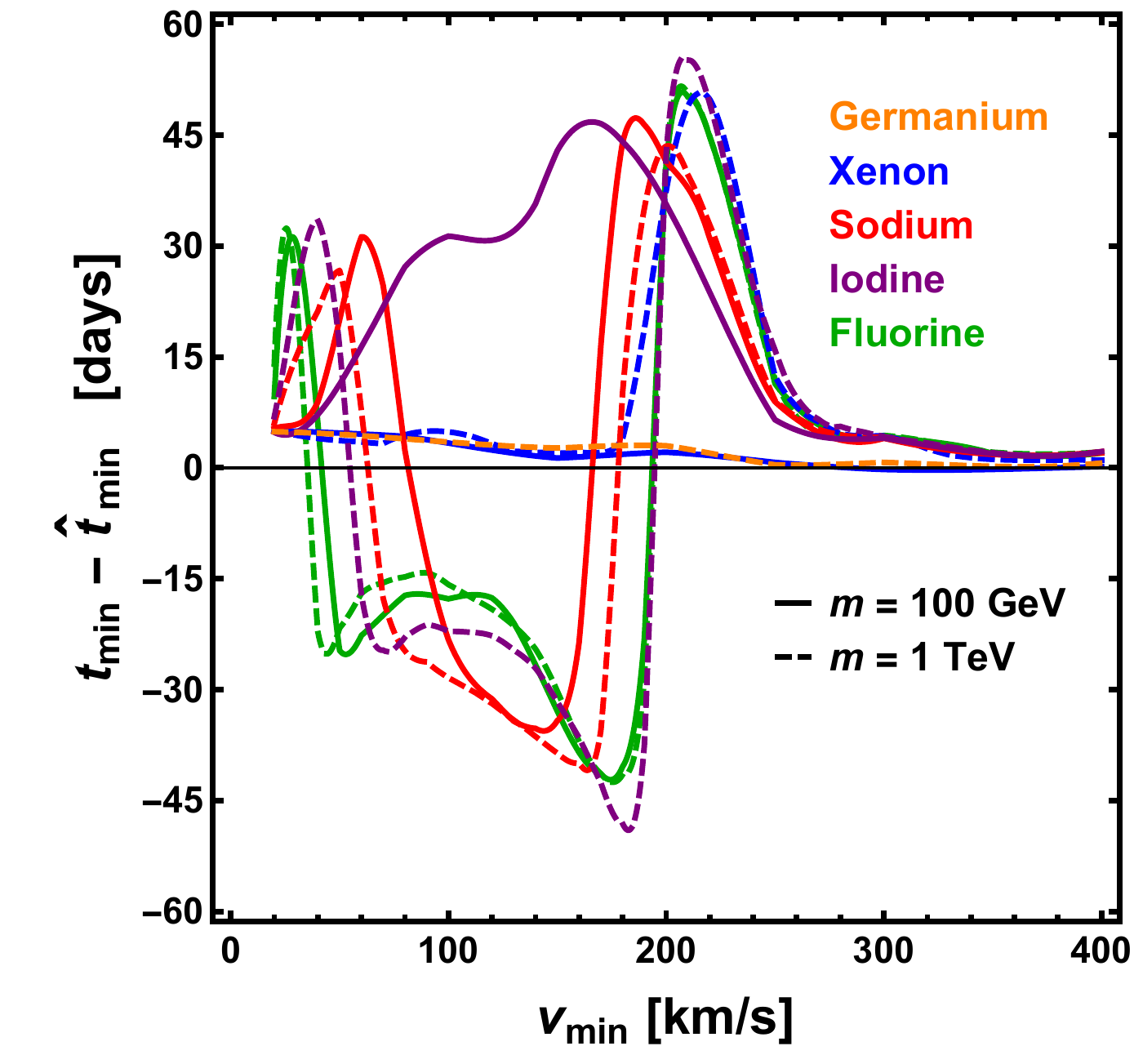}
\caption{\label{fig:magneticDMdiff} $\tmin - \htmin$, with $\htmin$ the time six month apart from $\tmax$. See \Fig{fig:magneticDM} for details.}
\end{figure}

We emphasize that the interplay between $\eta$ and $\tilde{\eta}$ does not only affect observables associated with the modulation of the rate, such as $\tmax$ and $\tmin$, but also the extent to which the standard approximation of the modulation given in \Eq{modrate} holds. \Fig{fig:magneticDMdiff} shows that the difference between $\tmin$ and $\htmin \equiv \tmax - 6 \text{ months}$ is target and DM particle mass dependent, and can be large, \eg $\tmin - \htmin$ for $m = 100$ GeV could be as large as $\pm 45$ days. This implies that higher order terms in the Fourier expansion of the rate beyond \Eq{modrate} cannot be neglected.

To illustrate how important the target element dependence of the rate modulation can be, consider the signal due to a $100$ GeV DM particle being detected with both xenon and fluorine near the present LUX and PICO thresholds. Were the modulation due solely to $\eta$ or $\tilde{\eta}$, the two experiments should observe nearly the same value of $\tmax$, see \Fig{fig:eta}. Instead, due to the target-dependent interplay of $\eta$ and $\tilde{\eta}$, the $\tmax$ observed with the two target elements could differ by more than four months and the modulation in xenon would be better described by \Eq{modrate} than the modulation in fluorine.

As we already mentioned, in order to observe the target-dependent effects described so far, it is essential that the experimental threshold in $\vmin$, which depends on the threshold in $\ER$, the DM particle mass and the scattering kinematics, is below $300$ km/s. \Fig{fig:magneticDM} shows that $m = 100$ GeV is already large enough with present thresholds to observe this effect. For lower $m$ the effect will only be present with the light targets, for elastic scattering.

Should DM scatter inelastically off nuclei, the scattering kinematics would be different from that of elastic scattering. Inelastic scattering \cite{TuckerSmith:2001hy, Graham:2010ca} can happen if there are at least two almost degenerate DM particles with masses $m$ and $m + \delta$ ($\delta \ll m$). If the particle with mass $m$ scatters into the $m + \delta$ particle, $\vmin = \left| (m_T \ER / \mu_T) + \delta \right| / \sqrt{2 m_T \ER}$. In particular, if $\delta < 0$ (exothermic scattering \cite{Graham:2010ca}), the $\vmin$ value corresponding to given $\ER$ and $m$ can be much smaller than in the case of elastic scattering.

All the effects we have described here rely on having a DM-nucleus differential cross section with a particular $v$ dependence. The issue remains of how such a cross section could be identified experimentally. We believe that this would require observing an annual modulation in at least two experiments with different target materials. If the velocity and target dependence in the differential cross section factorize, the observables associated with the modulation as functions of $\vmin$ would be independent of the target element, for any DM distribution. However, experiments do not measure their signal in $\vmin$, but in energy, and the values of $m$ and $\delta$ entering the $\ER$--$\vmin$ relation are not known a priori. This problem could be overcome by comparing observables of the modulation, like $\tmax$ and $\tmin$, of at least two experiments employing different target materials, and trying to find values of $m$ and $\delta$ that reconcile the differences between observed modulations as functions of $\vmin$. Should there exist no $\ER$--$\vmin$ relation that would make the modulations as functions of $\vmin$ compatible across experiments, one may infer the differential cross section contains a non-factorizable velocity and target dependence.

\section{Conclusions}
It is usually assumed that the modulation of the expected differential rate in direct DM detection experiments, expressed as a function of $\vmin$ (the minimum DM speed necessary to impart a certain recoil energy to a target nucleus), does not depend on the target. We have shown instead that experiments employing different target materials could observe an entirely different annual modulation of their differential rate as a function of $\vmin$. This would be a signature of DM interactions with more than one velocity-dependent term in the scattering cross section, in particular terms proportional to $1 / v^2$ and $v^n$ with $n \geqslant 0$. In order to identify experimentally this type of cross section, we believe at least two experiments employing different target materials should observe an annual modulation. Should no $\ER$--$\vmin$ relation be found that reconciles the modulated signals as functions of $\vmin$, one may infer the differential cross section contains a non-factorizable velocity and target dependence regardless of the DM distribution.

As an example, we have shown explicitly the target dependence of the time of maximum $\tmax$ and minimum $\tmin$ of the rate for a $100$ GeV and a $1$ TeV magnetic dipole DM scattering elastically assuming the SHM. We found that the values of $\tmax$ observed with \eg xenon and fluorine close to the present LUX and PICO thresholds could disagree by as much as four months (see \Fig{fig:magneticDM}), and the modulation in xenon could be better described by the sinusoidal time dependence usually assumed than that in fluorine.

\section{Acknowledgments} 
E.D.N.~and G.G.~acknowledge partial support from the Department of Energy under Award Number DE-SC0009937.

\bibliographystyle{apsrev4-1}
\bibliography{biblio}

\end{document}